\documentclass{mem}
\usepackage{natbib}\usepackage{txfonts}\usepackage{balance}
\usepackage{graphicx}
\usepackage[a4paper]{hyperref}
\idline{0}{1}
\begin{document}
\def\teff{$T\rm_{eff }$}
\def\kms{$\mathrm {km s}^{-1}$}

\title{
Pairing of Supermassive Black Holes\\in unequal-mass mergers
}

   \subtitle{}

\author{
S. \,Callegari\inst{1} 
\and L. \, Mayer\inst{1,2}
\and S. \, Kazantzidis\inst{3}
          }

  \offprints{S. Callegari}

\institute{
Institute for Theoretical Physics, University of Z\"urich,
Winterthurerstrasse 190, CH 8050,
Z\"urich, Switzerland,
\email{callegar@physik.uzh.ch}
\and
Institute of Astronomy, Department of Physics, 
ETH Z\"urich, Wolfgang-Pauli-Strasse 27, CH 8093,
Z\"urich, Switzerland
\and
Center for Cosmology and AstroParticle Physics,
Ohio State University,
191 West Woodruff Avenue, OH 43210,
Columbus, Ohio (USA)
}

\authorrunning{Callegari }

\titlerunning{Pairing of SMBHs in unequal-mass mergers}

\abstract{Disk galaxies with a spheroidal component are known to 
host Supermassive
  Black Holes (SMBHs) in their center. 
Unequal-mass galaxy mergers have been rarely studied despite the
fact that they are the large majority of merging events by number and
they are associated with the typical targets of gravitational wave
experiments such as LISA.
We perform N-body/SPH simulations of disk galaxy mergers with mass
ratios 1:4 and 1:10 at redshifts z=0 and z=3. They have the highest
resolution achieved so far for merging galaxies, and include star
formation and supernova feedback. Gas
dissipation is found to be necessary for the pairing of SMBHs in
these minor mergers. Still, 1:10 mergers with gas allow an efficient
pairing only at high z when orbital times are short enough compared to
the Hubble time.  
\keywords{black hole physics --- cosmology: theory --- galaxies: mergers
--- hydrodynamics --- methods: numerical}}
\maketitle{}

\section{Introduction}

Observations of nearby galaxies show that 
Supermassive Black Holes (SMBHs) ranging between $\sim10^6$ and
$\sim10^9 M_\odot$ inhabit the centers of virtually all massive
galactic spheroids, from massive ellipticals 
to pseudo-bulges of late-type galaxies. 
Their masses, as inferred from
dynamical measurements,  
appear to be correlated with various
properties of their hosts, e.g. bulge luminosity and mass
(e.g. \citet{kormendyrichstone95, magorrian98, marconihunt03}),
velocity dispersion  
\citep{tremaine02, ferrarese00},
concentration of the light profile \citep{graham01}.  
In the $\Lambda$CDM model, galaxies build up
their masses hierarchically starting from initial, gravitationally
amplified fluctuations (e.g. \citet{whiterees78}); 
therefore, every time two galaxies merge, the
remnant is expected to host two (or more) SMBHs. 
The formation of a binary SMBH sistem
has been shown to proceed quickly once both the compact
objects are embedded in a circumnuclear gaseous disk (see
\citet{lucio07}; see also 
Mayer, Kazantzidis \& Escala in this volume), but whether the 
large scale merger can lead them to such a favorable configuration is
still a matter of debate:
while mergers between galaxies of equal mass have been quite
extensively explored in literature and seem to lead to
the formation of a SMBH pair \citep{stelios05,springel05}, 
little attention has been paid to the fate of SMBHs
in events which are much more frequent in the typical history of a
$\Lambda$CDM galaxy, i.e. minor mergers with mass ratios from 1:4 to
1:10 and less \citep{stewart08}.

\section{Simulations and results}

\begin{figure*}[t!]
\resizebox{\hsize}{!}{\includegraphics[clip=true]{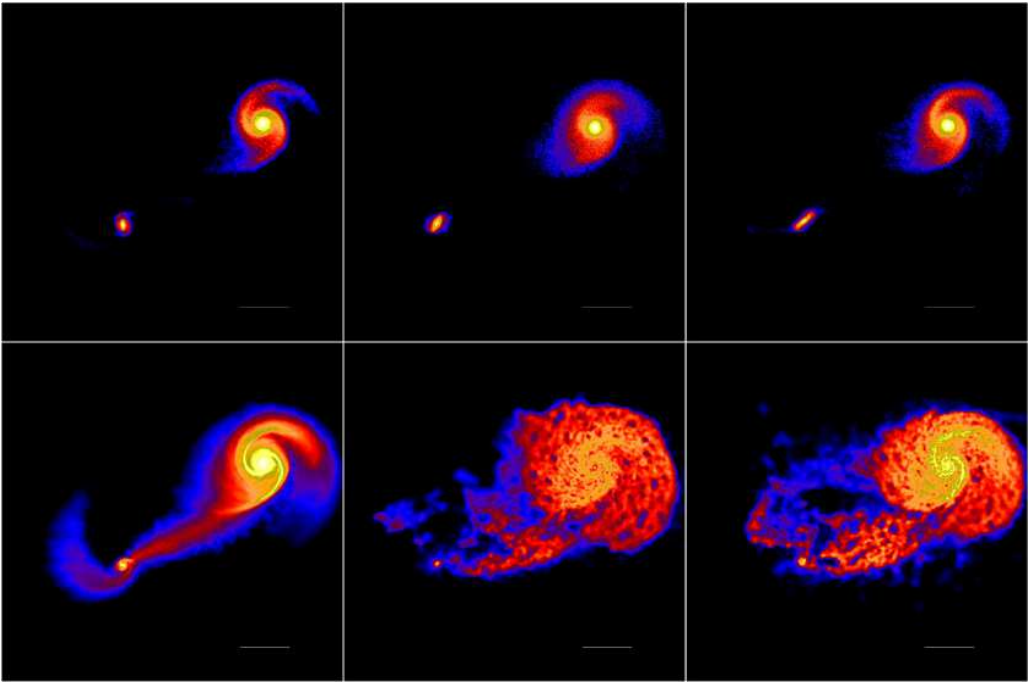}}
\caption{\footnotesize
Color-coded density maps for stars (upper row) and gas (lower row) at
the second apocenter for
three different $q=0.1$ mergers at $z=3$: from left to right, the case
with gas gooling ($f_g=10\%$), star formation ($f_g=10\%$) and star
formation ($f_g=30\%$) are shown. Each frame is $30$ kpc across.
In the runs with star formation, supernova feedback creates a
clumpy and diffuse ISM, resulting in strong ram pressure stripping
when the orbit of the satellite passes through the gaseous disk of the
primary galaxy.
}
\label{fig:tile}
\end{figure*}

\subsection{Initial Conditions}

We discuss here results from N-body/SPH simulations of mergers between
disk galaxies with mass ratios $q=$ 0.25 and 0.1 at an unprecedented
resolution. 
The galaxy models were initialized as three-component systems comprising a
Hernquist bulge, an exponential disk with a gaseous mass
fraction $f_g$, and a dark matter halo with an adiabatically
contracted NFW profile, plus a central collionless
particle representing the SMBH, whose mass was chosen following the
$M_{\rm BH} - M_{\rm bulge}$ relation \citep{magorrian98}. For more details on the
setup of the initial conditions for the reference model see
\citet{stelios05}. 
We also ran mergers with initial conditions rescaled to
$z=3$ according to the \citet{mmw98} (MMW) and \citet{bullock01}
models, and keeping 
$V_{\rm vir}$ fixed, as expected for the high-$z$
progenitors of our present-day models \citep{wechsler02, li07}. 
A large fraction 
of the gravitational wave signal from coalescences of SMBH binaries is
expected to come from this cosmic epoch \citep{sesa05}, if SMBH
pairing is efficient.
The satellite galaxies are initialized self-consistently with the
same three-component structure of the primary, and a mass in each
component scaled down by $q$.
The baricenters of the two galaxies in each run were initally placed
at a separation equal to the sum of their virial radii, on
parabolic orbits with pericentric distances equal to $0.2$ times the
virial radius of the more massive halo \citep{khochfar06, benson05}.
We ran collisionless (``{\it dry}'', with $f_g=0$) and gasdynamical
(``{\it wet}'', $f_g=0.1$ and $0.3$) simulations using GASOLINE
\citep{gasoline}.
We will define two SMBHs a ``pair'' if their relative orbit shrinks to
a separation equal to our force softening ($\sim 20$ pc); at these
distances, \citet{lucio07} have shown that sinking proceeds very
quickly and a SMBH binary can be formed in $\sim 1$ Myr.
Since our main purpose is to check which processes favour or inhibit
this pairing, we do not employ any additional prescription in
order to to keep them at the center of their galaxies' potential wells
or to facilitate their orbital sinking.

\subsection{Collisionless Mergers}
\label{ss:orbits}

The galaxy merger and the sinking of the lighter SMBH towards the
more massive one
can be roughly divided in three stages: during the first, the
orbit of the satellite decays because of dynamical friction on the
halo of the primary; the second encompasses the most crucial phase of
mass stripping in the denstest, baryon-dominated region where the fate
of the SMBH pair is decided; in the final stage
the remnant settles to its final dynamical state.  
 
For $q=0.25$, dynamical friction on the dark matter halo of the
primary is efficient, and the satellite galaxy sinks down 
to a few $\sim10$ kpc from the center. However, a collisionless system
is not able to dissipate the energy gained through tidal shocks at
pericentric passages \citep{gnedin99b,taffoni03}; thus, the satellite
is disrupted before its SMBH can reach the center of the remnant. No
pair is formed, and the smaller SMBH is left several kiloparsec away
from the other; at these distances its dynamical friction timescale
\citep{chandra43} is longer than a Hubble time, but this is also where
gas dynamics can affect the sinking (see \ref{ss:gasdyn}).

On the other hand, even after 10
Gyr the 1:10, $z=0$ dry merger is not able to enter the regime where 
the inclusion of gas could speed up the sinking,
because dynamical friction in the halo is too slow.
Mergers at $z=3$, instead, allow for an efficient sinking on much
shorter timescales. The MMW scalings predict, for a given
$V_{\rm vir}$, masses and radii 
that are a factor of $H(z=3)/H_0\sim1/5$ smaller; as a consequence,
orbital periods 
associated with orbits of same energy and pericenter (in units of 
$R_{\rm vir}$) are reduced by the same factor.
Therefore, the 1:10 collisionless merger at $z=3$ is completed in
$\sim2.5$ Gyr. Like for $q=0.25$, a wandering SMBH is left at $\sim
10$kpc from the center of the remnant. 

\subsection{Gas Dynamics}
\label{ss:gasdyn}

Gas dynamics can change the orbital evolution 
only during the second stage of the merger, when the satellite is
moving through the densest, baryon-dominated region of the parent
galaxy; this is when tidal shocks and ram pressure 
become so strong that gaseous dissipation can
significantly affect mass stripping around the 
smaller SMBH. 

For this reason, the dry and wet, $q=0.25$  mergers differ
only after the first couple of orbits ($t\sim6$~Gyr).
As already pointed out in \citet{stelios05}, gaseous dissipation is a
key element in the second stage of SMBH pairing in unequal-mass
mergers: for 1:4 mergers, the presence of gas is a necessary and sufficient
condition for the formation of a pair at the force resolution. The
pair is
embedded in a nuclear gaseous disk of radius $\sim 500$~pc, supported
mainly by rotation ($\sim 200$~km~s$^{-1}$), and comprising a mass of
$\sim 3\cdot10^9$~M$_\odot$.

In the 1:10, $z=3$ merger with cooling, 
the gas in the outer disk of the satellite 
gets tidally stripped; however, most of it, 
after the first couple of pericentric passages, is funnelled by tidal
torques to the center of the satellite, 
ending up in a $\sim 7\cdot10^7$~M$_\odot$
circumnuclear  structure. 
This central overdensity steepens the mass profile of the satellite,
allowing it to survive subsequent
tidal shocks until it gets dragged down to the nucleus of the remnant,
where a SMBH pair is formed. 
The gaseous
nuclear structure in the remnant 
is disk-like, of thickness equal to our softening length and 
supported by rotation
in the disk plane (with rotational velocity $\sim 180$~km~s$^{-1}$).

\subsection{Star Formation}

We discuss now preliminary results from the last set of simulations,
which include star formation and feedback from supernovae according to
the prescriptions and fiducial parameters detailed in
\citet{stinson06}. A comparative view of three $q=0.1$, $z=3$ mergers
is shown in figure \ref{fig:tile}. 

When star formation is included, the Interstellar Medium (ISM) in the
disks shows a multiphase and irregular structure, with larger
scalelengths both in the radial and vertical direction. For this
reason, the first pericentric passage of the satellite is not able to
excite strong, coherent inflows which can steepen the density
profile. Thus, during the second orbit, ram pressure exerted by the
ISM of the primary galaxy strips
most of the gas away from the satellite. The fraction of gas remaining
in its center can then either be consumed by star formation, or
stripped when the orbit reaches 
the densest regions of the primary: in this case, the small SMBH might
not be able to pair with the more massive one before tidal shocks
disrupt the core of the satellite.  

\section{Conclusions}

Understanding the formation of SMBH binaries is of fundamental importance 
for the search of gravitational waves as well as for all studies of
black hole demography and host galaxy coevolution.   
SMBH pairing in unequal-mass mergers depends
crucially on gasdynamical effect: satellites are disrupted too
quickly, leaving wandering SMBHs in all the collisionless cases we
studied. {\it The presence of gas seems to be necessary for the pairing of
SMBHs}. While it also appears sufficient for $q=0.25$, lower mass
objects are more heavily affected by feedback from star formation,
and their outcome could be very 
sensitive to the details of the physical processes involved.

\bibliographystyle{aa}

\bibliography{callegari}

\end{document}